\documentstyle[12pt,fleqn]{article}

\columnwidth\textwidth
\begin{document}

\newcommand{\mpi}{M^{2}_{\pi}}
\newcommand{\mk}{M^{2}_{K}}
\newcommand{\mv}{M_V}
\newcommand{\mvq}{M^2_V}
\newcommand{\mrho}{M_{\rho}}
\newcommand{\mrhoq}{M^2_{\rho}}
\newcommand{\mw}{M_{\omega}}
\newcommand{\mwq}{M^2_{\omega}}
\newcommand{\mphi}{M_{\phi}}
\newcommand{\mphiq}{M^2_{\phi}}
\newcommand{\mks}{M_{K^{\ast}}}
\newcommand{\mksq}{M^2_{K^{\ast}}}
\newcommand{\mh}{\hat{m}}
\newcommand{\mn}{\stackrel{\circ}{m}}

\newcommand{\fq}{F^{2}_0}
\newcommand{\fpi}{F_{\pi}}
\newcommand{\fpiq}{F^{2}_{\pi}}
\newcommand{\fv}{F_V}
\newcommand{\fvq}{F^2_V}
\newcommand{\frho}{F_{\rho}}
\newcommand{\fw}{F_{\omega}}

\newcommand{\Uy}{U^{\dagger}}
\newcommand{\uy}{u^{\dagger}}
\newcommand{\xy}{\chi^{\dagger}}
\newcommand{\vu}{V_{\mu\nu}}
\newcommand{\vo}{V^{\mu\nu}}
\newcommand{\vmu}{V_{\mu}}
\newcommand{\fmu}{F^{I}_{\mu}}
\newcommand{\fnu}{F^{I}_{\nu}}
\newcommand{\fr}{F_{R}^{\mu\nu}}
\newcommand{\fl}{F_{L}^{\mu\nu}}
\newcommand{\fmunu}{F_{R,L}^{\mu\nu}}
\newcommand{\w}{\omega}
\newcommand{\ro}{\rho^0}

\newcommand{\larw}{{\cal L}_{2}^{\rho\omega}}
\newcommand{\lav}{{\cal L}_{2}^V}
\newcommand{\la}{{\cal L}_{2}}
\newcommand{\mq}{{\cal M}}
\newcommand{\La}{{\cal L}}
\newcommand{\dmo}{d^{\mu}}
\newcommand{\du}{d_{\mu}}
\newcommand{\pa}{\partial}
\newcommand{\pu}{\partial_{\mu}}
\newcommand{\po}{\partial^{\mu}}

\newcommand{\beq}{\begin{equation}}
\newcommand{\beqn}{\begin{eqnarray}}
\newcommand{\eeq}{\end{equation}}
\newcommand{\eeqn}{\end{eqnarray}}
\newcommand{\no}{\nonumber}
\newcommand{\nol}{\nonumber\\}

\pagestyle{empty}
\pagenumbering{arabic}

\begin{flushright}
TTP 95-08\\
April 1995\\[15mm]
\end{flushright}

\begin{center}
{\Large {\bf $\rho^0 -\omega$ mixing in chiral perturbation theory}}\\[15mm]
Res Urech\\[2mm]
Institut f\"ur Theoretische Teilchenphysik\\
Universit\"at Karlsruhe\\
D-76128 Karlsruhe, Germany\\[4mm]
e-mail: ru@ttpux2.physik.uni-karlsruhe.de\\[18mm]
{\large {\bf Abstract}}\\
\end{center}
In order to calculate the $\rho^0 -\omega$ mixing we extend the chiral
couplings of the low-lying vector mesons
in chiral perturbation theory to a lagrangian that
contains two vector fields. We determine the $p^2$ dependence of the
two-point function and recover an earlier result for the on-shell expression.
We discuss the off-shell behaviour of the mixing.\\[5mm]

\section{Introduction}

Chiral perturbation theory (CHPT) \cite{we79,ga84,ga85} is an effective low
energy theory that describes
first of all the interactions of the pseudoscalar meson octet. They are the
Goldstone bosons of the spontaneously broken chiral symmetry in QCD with
three light flavours. The chiral lagrangian is expanded perturbatively in
the three small quark masses and in the derivatives of the fields. At
lowest order $p^2$ the effective lagrangian is given by the non-linear
$\sigma$-model coupled to external fields.
In \cite{ga84} the authors include the $\rho$ meson in order to estimate
the next to leading order coupling constants in CHPT via resonance exchange,
in \cite{eck89} this work is extended to general meson exchange in chiral
$SU(3)$. The couplings of the resonances to the pseudoscalar mesons are
considered at lowest order in the chiral counting and linear in the
resonance fields.\newline
In the vector meson sector, $(\rho^{+} ,\ro ,\rho^{-} )$ represents a
(nearly) degenerated isospin triplet, whereas $\w$ is an isospin singlet.
However, isospin is not an exact symmetry, it is broken by the quark mass
difference $m_u -m_d$ and the electromagnetic interaction and therefore
$\ro$ and $\w$ mix. We calculate this mixing in the framework
of CHPT.\\[2mm]
Here we determine the lowest order couplings of the pseudoscalar mesons
to the vector mesons quadratic in the vector fields. We restrict
ourselves to the couplings that we need for $\ro -\w$ mixing and
represent the vector fields in the antisymmetric tensor notation
\cite{ga84,eck89}. It has been proven \cite{eck289} that this
representation is equivalent to the vector field formulation, to the
description that involves massive Yang-Mills fields \cite{mei88} and to the
model
with hidden gauge vector bosons \cite{ban88}.\newline
We give the two-point function for $\ro -\w$ mixing and find that the
on-shell expression corresponds to the result given by Gasser and
Leutwyler \cite{ga82}. Numerically it can be estimated from the decay $\w
\rightarrow \ro \rightarrow\pi^+\pi^-$. The mixing is
dominated by the strong interaction,
determined at this order by  the quark mass ratio $R=(m_s -\mh)/(m_d
-m_u)$ with $\mh=(m_u+m_d)/2$. The branching ratio
$B(\w \rightarrow \pi^+\pi^-)$ has changed since 1982 from 1.4 to 2.2\%,
therefore we give an update of the value of $R$.\newline
The vector mesons in the tensor representation can be related to the usual
vector formalism by the definition (see section 2)
\beq
V_\mu = \frac{1}{M}\pa^\nu V_{\nu\mu},
\eeq
where $M$ denotes the mass of the particle.
With this definition, the  couplings of the vector mesons are soft: for a
vanishing four-momentum the vector resonances
decouple. This implies that the $\ro -\w$ mixing amplitude has a zero at
$p^2=0$. O'Connell
et al. \cite {th94} have shown that a broad class of models contains this
feature. However, the decoupling which occurs in two-point functions and in
hadronic
form factors \cite{cal76} does not a priori show up in matrix elements
without external vector mesons,
e.g. nucleon-nucleon scattering with resonance exchange. This point will be
discussed in section 4.\newline
In section 2 we present the lagrangian of CHPT at leading order and give the
properties of the vector mesons in the antisymmetric tensor notation. In
section 3 we determine the couplings quadratic in the resonance fields and
pin down the coupling constants by using large $N_C$ arguments \cite{wit78},
the Zweig rule \cite{zwei65}, and the quark model
counting for the vector meson masses at first order in the quark mass
expansion \cite{ga82}.
In section 4 we calculate the two-point function for $\ro -\w$ mixing and
compare our numerical result for the on-shell expression with the
calculations found in the literature. We evaluate the quark mass ratio $R$
{}from the mixing amplitude and discuss the off-shell behaviour of the
two-point
function and its consequence on nucleon-nucleon scattering involving
$\ro-\w$  exchange.\newline
For a introduction to the various topics we refer the reader to recent
reviews  of CHPT \cite{chpt} and $\ro-\w$ mixing \cite{rho,th95}.
\pagestyle{plain}

\section{Field properties and dynamics}

The chiral lagrangian can be expanded in derivatives of the Goldstone fields
and in the masses of the three light quarks. One derivative counts as
a quantity of $O(p)$, the masses are of order $O(p^2)$ and the fields
themselves are $O(p^0)$. The effective lagrangian starts at $O(p^2)$,
denoted by $\la$. It is the non-linear $\sigma$-model lagrangian coupled to
external fields, respects chiral symmetry $SU(3)_{R}\times
SU(3)_{L}$ and is invariant under $P$ and $C$ transformations
\cite{ga84,ga85},
\beq
\la=\frac{1}{4}\fq\langle\dmo\Uy\du U+\chi\Uy +\xy U\rangle .
\eeq
The brackets $\langle\cdots\rangle$ denote the trace in flavour space, $U$
is a unitary $3\times 3$ matrix and incorporates the fields of the
eight pseudoscalar mesons,
\beqn
U&=&exp\,(i\Phi/F_0),\hspace{10mm}\det U=1,\nol[4mm]
\Phi&=&\sqrt{2}\left(
\begin{array}{ccc}
\frac{1}{\sqrt{2}}\pi^0+\frac{1}{\sqrt{6}}\eta_8 & \pi^{+} & K^{+}\\
\pi^{-} & -\frac{1}{\sqrt{2}}\pi^0+\frac{1}{\sqrt{6}}\eta_8 & K^0\\
K^{-} & \bar{K}^0 & -\frac{2}{\sqrt{6}}\eta_8
\end{array}
\right).
\eeqn
$\du U$ is a covariant derivative incorporating the external vector and
axialvector currents $v_{\mu}$ and $a_{\mu}$, respectively,
\beq
\du U = \pu U - i(v_{\mu} + a_{\mu}) U + i U (v_{\mu} - a_{\mu}),
\eeq
disregarding singlet (axial)vector currents, we put
${\rm tr}\;v_{\mu}={\rm tr}\;a_{\mu}=0$.
$\chi$ represents the coupling
of the mesons to the scalar and pseudoscalar  currents $s$ and $p$,
respectively, and $s$ incorporates the mass matrix
\beqn
\chi&=&2B_0(s+ip),\nol
s&=&\mq+\cdots=\left(
\begin{array}{ccc}
m_{u}&&\\
&m_{d}&\\
&&m_{s}
\end{array}\right)+\cdots
\eeqn
$F_0$ is the pion decay constant in the chiral limit $(\mq =0)$,
$\fpi=F_0 \left[1+O(m_{q})\right]$ and $B_0$ is related to the quark
condensate
$\langle 0|\bar{u}u|0\rangle =-\fq B_0 \left[1+ O(m_{q})\right]$.
The transformation properties under $SU(3)_{R}\times SU(3)_{L}$ are
\beqn
U&\rightarrow& g_{R}Ug^{\dagger}_{L}\nol
v_{\mu} + a_{\mu}&\rightarrow& g_{R}(v_{\mu} +
a_{\mu}) g^{\dagger}_{R} + ig_{R}\pu g^{\dagger}_{R}\nol
v_{\mu} - a_{\mu}&\rightarrow& g_{L}(v_{\mu} -
a_{\mu})g^{\dagger}_{L} +ig_{L}\pu g^{\dagger}_{L}\\
s+ip&\rightarrow& g_{R}(s+ip)g^{\dagger}_{L}\nol
g_{R,L}&\in& SU(3)_{R,L}.\no
\eeqn
To be consistent in the chiral counting, $v_{\mu}$ and $a_{\mu}$ count as
O(p),  $s$ and $p$ are of order $O(p^2)$.
For the representation of the vector mesons we follow
\cite{ga84,eck89,eck289}, using the antisymmetric tensor notation
$V_{\mu\nu}$. The lagrangian for free particles is given by
\beq
\La_{free}^V=-\frac{1}{2}\po V_{\mu\nu} \pa_{\rho}V^{\rho\nu} +\frac{1}{4}
M^2 V_{\mu\nu}V^{\mu\nu},
\eeq
{}from where we derive the equation of motion,
\beq\label{eqom}
\po \pa_{\rho}V^{\rho\nu} - \pa^{\nu}\pa_{\rho}V^{\rho\mu} + M^2V^{\mu\nu}=0,
\eeq
and the free propagator
\beqn\label{twop}
\lefteqn{\langle 0|T\,V_{\mu\nu}(x)V_{\rho\sigma}(y)|0\rangle =i\int
\frac{d^4 k}{(2\pi)^4}e^{-ik(x-y)}\Delta_{\mu\nu\rho\sigma},}\nol
\Delta_{\mu\nu\rho\sigma}&=&\frac{1}{M^2}\left(G_{\mu\nu\rho\sigma}
+ \frac{1}{M^2 - k^2}P_{\mu\nu\rho\sigma}\right)\\
G_{\mu\nu\rho\sigma}&=&g_{\mu\rho}g_{\nu\sigma} -
g_{\mu\sigma}g_{\nu\rho}\nol
P_{\mu\nu\rho\sigma}&=&g_{\mu\rho}k_\nu k_\sigma -
g_{\mu\sigma}k_\nu k_\rho - g_{\nu\rho}k_\mu k_\sigma
+g_{\nu\sigma}k_\mu k_\rho .\no
\eeqn
This corresponds to the normalization $\langle 0|V_{\mu\nu}|
V(p)\rangle=i(p_\mu\epsilon_\nu -p_\nu\epsilon_\mu)/M$, where
$\epsilon_\mu$ represents the polarization vector. With the definition
\beq\label{vector}
V_\mu = \frac{1}{M}\pa^\nu V_{\nu\mu}
\eeq
we obtain from (\ref{eqom}) the Proca equation, $\pa_\rho (\pa^\rho
V^\mu - \po V^\rho ) + M^2 V^\mu =0$, and the two-point function
\beq
\langle 0|T\,V_\mu (x)V_\nu (y)|0\rangle =i\int \frac{d^4 k}{(2\pi)^4}
e^{-ik(x-y)}\left(g_{\mu\nu} - \frac{k_\mu
k_\nu}{M^2}\right)\frac{1}{M^2 -k^2}.
\eeq
The normalization is given by $\langle 0|V_{\mu}|V(p)\rangle
=\epsilon_\mu$.\\[2mm]
The fields $V_{\mu\nu}$ carry non-linear realizations of $SU(3)$, the
transformation under the chiral group  $G=SU(3)_R\times SU(3)_L$ is
\cite{eck89}
\beqn
V_{\mu\nu}&\stackrel{G}{\longrightarrow}& h(\varphi)
V_{\mu\nu}h^\dagger(\varphi)
\hspace{2cm}{\rm octet},\nol
\w_{1\,\mu\nu}&\stackrel{G}{\longrightarrow}& \w_{1\,\mu\nu}
\hspace{3.6cm}{\rm singlet}.
\eeqn
The non-linear realization $h(\varphi)$ is defined by determining the action
of $G$ on an element $u(\varphi)$ of the coset space $\left[SU(3)_R\times
SU(3)_L\right]/SU(3)_V$ \cite{col69}
\beq
u(\varphi)\stackrel{G}{\longrightarrow} g_R u(\varphi)h^\dagger(\varphi)=
h(\varphi)u(\varphi)g_L^\dagger,
\eeq
where $\varphi$ represents the Goldstone bosons and $U=u^2(\varphi)$. For
the octet part, a covariant derivative is defined as
\beq
\nabla^{\rho}V_{\rho\mu}= \pa^{\rho} V_{\rho\mu} +
[\Gamma^{\rho},V_{\rho\mu}]
\eeq
with
\beq
\Gamma^{\rho}= \frac{1}{2}\left\{\uy[\pa^{\rho} -i(v^{\rho}+a^{\rho})]u +
u[\pa^{\rho} -i(v^{\rho}-a^{\rho})]\uy\right\}
\eeq
and $V_{\mu\nu}$ in the matrix notation
\beq
V_{\mu\nu}=\left(
\begin{array}{ccc}
\frac{1}{\sqrt{2}}\ro_{\mu\nu}+\frac{1}{\sqrt{6}}\w_{8\,\mu\nu} &
\rho^{+}_{\mu\nu} & K^{\ast\,+}_{\mu\nu}\\[2mm]
\rho^{-}_{\mu\nu} &
-\frac{1}{\sqrt{2}}\ro_{\mu\nu}+\frac{1}{\sqrt{6}}\w_{8\,\mu\nu} &
K^{\ast\,0}_{\mu\nu} \\[2mm]
K^{\ast\,-}_{\mu\nu} & \bar{K}^{\ast\,0}_{\mu\nu} &
-\frac{2}{\sqrt{6}}\w_{8\,\mu\nu}
\end{array}
\right).
\eeq
The covariant derivative is transformed like the field itself
\beq
\nabla^{\rho}V_{\rho\mu}\stackrel{G}{\longrightarrow} h(\varphi)
\nabla^{\rho}V_{\rho\mu}h^\dagger(\varphi).
\eeq
Denoting the octet (singlet) mass in the chiral limit by $\mv$ ($M_{\w_1}$),
the kinetic part of the lagrangian takes the form
\beqn
\La_{kin}^V&=&-\frac{1}{2}\langle \nabla^{\rho}V_{\rho\mu}
\nabla_{\sigma}V^{\sigma\mu} - \frac{1}{2}\mvq
V_{\mu\nu}V^{\mu\nu}\rangle\nol
&&-\frac{1}{2}\pa^{\rho} \w_{1\,\rho\mu}\, \pa_{\sigma} \w_1^{\sigma\mu}
+\frac{1}{4}M^2_{\w_1} \w_{1\,\mu\nu}\w_1^{\mu\nu}.
\eeqn
The interaction of the pseudoscalar mesons with one vector meson
starts at order $O(p^2)$. The interaction lagrangian contains the octet
fields only, there is no coupling to the singlet at this order \cite{eck89},
\beq\label{lv}
\lav=\frac{\fv}{2\sqrt{2}}\langle V_{\mu\nu}f_+^{\mu\nu}\rangle
+\frac{iG_V}{2\sqrt{2}}\langle V_{\mu\nu}[u^{\mu},u^{\nu}]\rangle,
\eeq
where
\beqn
f_+^{\mu\nu}&=&u\fl\uy + \uy\fr u\nol
\fmunu&=&\po (v^\nu\pm a^\nu) -\pa^\nu(v^\mu\pm a^\mu)-i[v^\mu\pm a^\mu,
v^\nu\pm a^\nu]\\
u^\mu&=&i\uy\dmo U\uy=u^{\dagger\,\mu}.\no
\eeqn
We refer to \cite{eck89} for a complete list of all the terms that can
couple to the resonances and their properties under $P$ and $C$
transformations. $\fv$ and $G_V$ are real coupling constants that are not
restricted by chiral symmetry \cite{eck289}. For later use we include the
electromagnetic interactions in $\lav$ by adding the photon field
$A_\mu$ to the vector current,
\beq
v_\mu\pm a_\mu \rightarrow v_\mu +QA_\mu\pm a_\mu,
\eeq
where $Q$ is the charge matrix of the three light quarks, $Q=e\;
{\rm diag}(2/3,-1/3,-1/3)$. The corresponding kinetic part of the photons
reads
\beq
\La_{kin}^\gamma = -\frac{1}{4} F_{\mu\nu} F^{\mu\nu} - \frac{1}{2} \left(
\pu A^{\mu} \right)^{2}
\eeq
with  $F_{\mu\nu}=\pu A_{\nu}-\pa_\nu A_{\mu}$ and the gauge fixing
parameter chosen to be $\lambda =1$.

\section{Interaction quadratic in the vector mesons}

In this section we construct the lagrangian with two vector fields coupled
to the pseudoscalar mesons. We restrict ourselves to the part
that is relevant for $\ro -\w$ mixing at lowest order,
\beqn
\larw&=&v_8\langle V_{\mu\nu}V^{\mu\nu}\chi_+\rangle +
\tilde{v}_8 \langle V_{\mu\nu}V^{\mu\nu}\rangle\langle\chi_+\rangle\nol
&&+v_{18}\,\w_{1\,\mu\nu}\langle V^{\mu\nu}\chi_+\rangle
+v_1\, \w_{1\,\mu\nu}\,\w_1^{\mu\nu}\langle\chi_+\rangle ,
\eeqn
where $\chi_+=\uy\chi\uy + u\xy u$.
In order to pin down the coupling constants in $\larw$ we make the
following assumptions.\newline
$(i)$ In the large $N_C$ limit \cite{wit78} the product of two traces is
suppressed by at least one power of $N_C$. Therefore we will put
$\tilde{v}_8=0$. Furthermore, for $N_C\rightarrow \infty$ the octet and
singlet mesons are degenerated and thus $\mv=M_{\w_1}$.\newline
$(ii)$ The Zweig rule \cite{zwei65} implies that the $\w$-meson is free of
$s\bar{s}$, the $\phi-\w$ mixing is ideal,
\beqn
|\phi\rangle&=&\cos\theta_V|\w_8\rangle - \sin\theta_V|\w_1\rangle,\nol
|\w\rangle&=&\sin\theta_V|\w_8\rangle + \cos\theta_V|\w_1\rangle
\eeqn
with $\tan\theta_V=1/\sqrt{2}$. On the same footing we assume that
there is no $\ro-\phi$ mixing in the lagrangian $\larw$.\newline
$(iii)$ For the masses of the vector mesons at order $O(p^2)$ we invoke the
quark counting rule \cite{ga82}. We keep the eigenstates of the kinetic
lagrangian $\La_{kin}^V$, there are no first order mass shifts from $\ro
-\w$ mixing. In addition we work (for the masses only) in the isospin limit
$\mh=m_u=m_d$,
\beqn\label{mass}
\mrho&=&\mw =\mv +2\mh,\nol
\mks&=&\mv +\mh +m_s,\\
\mphi&=&\mv +2m_s.\no
\eeqn
The nearly flavour independent  differences  \cite{pdg}
\beq
\mrhoq - \mpi = 0.58\;{\rm GeV}^2,
\hspace{1cm}\mksq - \mk = 0.55\;{\rm GeV}^2,
\eeq
and the lowest order mass formulae for the Goldstone bosons
\beq
\mpi =2B_0\mh,\hspace{1cm}\mk = B_0(\mh + m_s),
\eeq
lead to the relation $\mv\simeq B_0/2$.
Taking all the assumptions into account, we arrive at
\beq
v_8=\frac{1}{8},\hspace{1cm}\tilde{v}_8=0,\hspace{1cm}
v_{18}=\frac{1}{4\sqrt{3}},\hspace{1cm}v_1=\frac{1}{24}.
\eeq
The lagrangian takes the simple form (omitting terms that are not
relevant for $\ro -\w$ mixing)
\beqn\label{larw}
\La_{kin}^V + \larw&=&-\frac{1}{2}\pa^{\lambda} \ro_{\lambda\mu}\,
\pa_{\sigma}
\rho^{0\,\sigma\mu} +\frac{1}{4}\mrhoq \ro_{\mu\nu}\rho^{0\,\mu\nu}\nol
&&-\frac{1}{2}\pa^{\lambda} \w_{\lambda\mu}\, \pa_{\sigma}
\w^{\sigma\mu} +\frac{1}{4}\mwq\,\w_{\mu\nu}\w^{\mu\nu}
+\mrho(m_u -m_d) \ro_{\mu\nu}\w^{\mu\nu},
\eeqn
where we have replaced $\mv$ by $\mrho$ in the interaction term, but kept
in the mass term the different notation for $\mw$ and $\mrho$.

\section{Mixing amplitude}

The contributions to lowest order $\ro-\w$ mixing are the contact term in
(\ref{larw}) and the one-photon exchange from the lagrangian $\lav$ linear
in the
vector fields in (\ref{lv}). The Fourier transform of the two-point
function in the tensor notation has the form
\beqn\label{mixtwo}
\lefteqn{i\int d^4xe^{ipx}
\langle 0|T\,\ro_{\mu\nu}(x)\w_{\rho\sigma}(0) e^{i\int d^4y\left\{\lav
+\larw\right\}} |0\rangle =}\nol[2mm]
&&\frac{2\mrho(m_u -m_d)}{\mrhoq\mwq}\left\{
G_{\mu\nu\rho\sigma} +\left[\frac{1}{\mrhoq -p^2} +\frac{\mrhoq}{(\mrhoq
-p^2)(\mwq -p^2)}\right]P_{\mu\nu\rho\sigma}\right\}\nol[2mm]
&&+\frac{1}{3}\frac{e^2\fvq}{(\mrhoq -p^2)(\mwq -p^2)p^2}
P_{\mu\nu\rho\sigma}.
\eeqn
For the calculation of the on-shell amplitude, we consider the decay
$\w\rightarrow\ro\rightarrow\pi^+\pi^-$. We find
\beqn\label{wdecay}
\Gamma(\w\rightarrow \pi^+\pi^-)&=&\frac{\Theta^2_{\rho\w}
\Gamma(\ro\rightarrow \pi^+\pi^-)}{|\mwq -\mrhoq -i(\mw\Gamma_{\w}
-\mrho\Gamma_{\rho})|^2}\nol
&\simeq&\frac{\Theta^2_{\rho\w}}{4\mrhoq}\,\cdot\,
\frac{\Gamma(\ro\rightarrow
\pi^+\pi^-)}{(\mw -\mrho)^2 +\frac{1}{4}(\Gamma_{\w}
-\Gamma_{\rho})^2},
\eeqn
where we introduced the widths $\Gamma_\rho$ and $\Gamma_\w$ in the
propagator and
\beqn\label{thamp}
\Gamma(\ro\rightarrow \pi^+\pi^-)&=&\frac{1}{48\pi}
\frac{G_V^2\mrho^3}{F^4_0} \left(1-\frac{4\mpi}{\mrhoq}\right)^{3/2},\nol
\Theta_{\rho\w}&=&2\mrho(m_u-m_d)+\frac{1}{3}e^2\fvq.
\eeqn
The amplitude $\Theta_{\rho\w}$ has already been determined in \cite{ga82},
where
the strong interaction part was derived in a quantum mechanical approach,
\beq\label{thga}
\Theta_{\rho\w}=2\mrho\left[-\frac{m_d-m_u}{m_s-\mh}(\mks-\mrho)
+e^2\frac{\frho\fw}{2\mrho}\right].
\eeq
If we identify $\frho=\fv$, $\fw=\frac{1}{3}\fv$ and insert the masses for
the vector mesons from (\ref{mass}) we get the same expression for
$\Theta_{\rho\w}$. With the experimental decay widths \cite{pdg} we find the
value
\beq\label{thnum}
\Theta_{\rho\w}=(-3.91\pm 0.30)\times 10^{-3}\;{\rm GeV}^2.
\eeq
The sign is determined from the relative phase of the $\w$ and $\rho$
amplitudes in $e^+e^-\rightarrow\pi^+\pi^-$ near $\mrho$ \cite{coon77}. The
error in (\ref{thnum}) is entirely due to the uncertainty in the decay width
$\Gamma(\w\rightarrow \pi^+\pi^-)$. If we neglect the mass difference $(\mw
-\mrho)$ and the width $\Gamma_\w$ in (\ref{wdecay}), the mixing amplitude
decreases to $\Theta_{\rho\w}=-4.08\times 10^{-3}\;{\rm GeV}^2$. Our result
in (\ref{thnum}) is in agreement with the values quoted in the
literature, with the exception of the recent result of Friedrich
and Reinhardt \cite{frie95},
\beq
\begin{array}{|c|c l|}
\hline
&&\\[-2mm]
\hspace{5mm}\Theta_{\rho\w} \left[\times 10^{-3}\;{\rm
GeV}^2\right]\hspace{5mm} &\hspace{5mm}& \hspace{12mm}{\rm
references}\\[2mm]
\hline
&&\\[-2mm]
- 4.52\pm 0.60 && \mbox {Coon and
Barrett}\hspace{2mm}\cite{coon87}\\[2mm]
\left.
\begin{array}{c}
\hspace{3mm}- 3.74\pm 0.30 \\
\hspace{3mm}- 4.23\pm 0.68 \\
\hspace{3mm}- 3.67\pm 0.30 \\
\end{array}
\right\}
&& \mbox{Bernicha et al.}\hspace{2mm}\cite{bern94}\\[8mm]
- 3.97\pm 0.20 && \mbox{O'Connell et al.}\hspace{2mm}\cite{th95}\\[2mm]
- 4.85 && \mbox{Friedrich and
Reinhardt}\hspace{2mm}\cite{frie95}\hspace{5mm}\\[2mm]
\hline
\end{array}
\eeq
{}From the expression in (\ref{thga}), Gasser and Leutwyler determined the
quark mass ratio $R=(m_s-\mh)/(m_d -m_u)$ to $R=44\pm 3$ \cite{ga82}.
The branching ratio $B(\w \rightarrow \pi^+\pi^-)$ changed in the meantime
{}from 1.4 to 2.2\%, motivation enough to give an update of $R$. The value
for $\frho$ can be obtained from the decay $\ro\rightarrow
e^+e^-$,
\beqn
\Gamma(\ro\rightarrow e^+e^-)&=& \frac{1}{12\pi}\frac{e^4\frho^2}{\mrho}
,\nol
\frho&=&152\pm 4\;{\rm  MeV}.
\eeqn
This leads to the estimate
\beq\label{quark}
R= 40.7\pm 3.0,
\eeq
where we assumed again that $\frho/\fw=3$, as it turns out from the
lagrangian $\lav$ in (\ref{lv}). Experimentally, this ratio is given by
\cite{pdg}
\beq\label{frho}
\frac{\frho}{\fw}=\left[ \frac{\mrho\Gamma(\ro\rightarrow
e^+e^-)}{\mw\Gamma(\w\rightarrow e^+e^-)}\right]^{1/2} =3.31\pm 0.15,
\eeq
which changes the meanvalue in (\ref{quark}) to $R=41.3$. At this stage,
there is poor information about the uncertainties in the contribution from
the strong interaction part. We can ask whether or not the assumption in
(\ref{mass}) is still a good approximation beyond leading order in
the quark mass expansion. For comparison we consider the kaon and nucleon
masses. One finds at the one loop level \cite{ga85,ger89,buer93}
\beqn\label{devmass}
\frac{\partial M_K}{\partial\mh}=\frac{\partial M_K}{\partial m_s}
&=&\frac{M_K}{2(\mh+m_s)}(1\pm 0.1),\nol
\mh\frac{\partial M_N}{\partial\mh}=\sigma_{\pi N}
&=&\sigma_{\pi N}^{tree}\left(1+\frac{6.8\;{\rm MeV} +0.9\times
10^{-2}\mn}{\sigma_{\pi N}^{tree}}\right).
\eeqn
The second entry in the bracket shows the correction from the one-loop
contribution. For $\sigma_{\pi N}$ we used a linear approximation (in
$\mn$) of
the result given in \cite{buer93}. $\mn$ denotes the mass of
the nucleon in the chiral limit and $\sigma_{\pi N}^{tree}=(92.7\;{\rm MeV}
-7.4\times 10^{-2}\mn)$. In contrast to the kaon mass, the one-loop
contribution to the $\sigma$ term is not small, it is between
$32\%(\mn\; =700\;{\rm MeV})$ and $57\%(900\;{\rm MeV})$.
We conclude that for the vector mesons the mass formula (\ref{mass})
{\it could} underestimate the valence quark content at next to leading order
by an overall factor $1.5\,$. The meanvalue of the quark mass ratio  would
change to  $R\sim 27\,$. We
leave this guess as it stands, we only take the electromagnetic deviations
{}from (\ref{frho}) into account, but keep in mind that there is a strong
dependence on the assumption in (\ref{mass}). We correct the value for $R$
by increasing the error bar and our estimate finally reads
\beq
R=41\pm 4.
\eeq
For the discussion of the off-shell mixing we consider again the two-point
function in (\ref{mixtwo}). With the definition in (\ref{vector}) we
arrive at
\beqn\label{twovec}
\lefteqn{i\int d^4xe^{ipx}
\langle 0|T\,\ro_{\mu}(x)\w_{\nu}(0) e^{i\int d^4y\left\{\lav
+\larw\right\}}|0\rangle =}\nol
&&\left(g_{\mu\nu} - \frac{p_\mu p_\nu}{p^2}\right)
\frac{\Theta(p^2)}{(\mrho^2 -p^2)(\mw^2 -p^2)},\nol
\Theta(p^2)&=&\frac{p^2}{\mrho\mw}\Theta_{\rho\w},
\eeqn
where $\Theta_{\rho\w}$ is given in (\ref{thamp}). $\Theta(p^2)$  contains
a zero at $p^2=0$ and is positive in the spacelike region. It has been
shown by O'Connell et al. \cite {th94} that in a wide range of models the
$\ro-\w$ mixing amplitude must vanish at $p^2=0$. Still we
cannot conclude that matrix elements involving $\ro$ and $\w$ as
intermediate states show the same decoupling. Consider nucleon-nucleon
scattering, where the baryons are again described by a $3\times 3$ matrix
that transforms under $SU(3)_R\times SU(3)_L$ in the same way as the meson
resonances \cite{krau90},
\beq
B\stackrel{G}{\longrightarrow} h(\varphi) B\,h^\dagger(\varphi).
\eeq
Chiral symmetry and the invariance under $P$ and $C$ transformations lead to
the interaction lagrangian $\La_0^{B\bar{B}V}$ at lowest order with two
baryons $B,\bar{B}=B^{\dagger}\gamma^0$ and one vector meson,
\beq
\La_0^{B\bar{B}V}=a_1\sigma^{\mu\nu}\langle \bar{B}\{V_{\mu\nu},B\}\rangle
+a_2\sigma^{\mu\nu}\w_{1\,\mu\nu}\langle \bar{B}B\rangle,
\eeq
where we leave the coupling constants $a_1,a_2$ undetermined, since we are
interested in the qualitative picture only and neglect the mass
difference $(\mrho-\mw)$ and the electromagnetic contribution in the
following. The amplitude of nucleon-nucleon scattering with
resonance exchange and $\ro-\w$ mixing from (\ref{mixtwo}) has the form
(see \cite{krau90} for conventions)
\beq\label{scatt}
T^{\rho\w}_{NN}(t)\sim a_{NN}\left(2 +\frac{t(2\mrhoq-t)}{(\mrhoq
-t)^2}\right)
=a_{NN}\frac{(\mrhoq -t)^2+\mrho^4}{(\mrhoq -t)^2},
\eeq
with $a_{NN}$ a flavour dependent constant (including $\Theta_{\rho\w}$)
and $t$ the momentum transfer.
We see a $t$ dependence in the numerator which reaches its minimum at
$t=\mrhoq$ and is positive everywhere, therefore it does not change the
sign in the spacelike region. Though the mixing amplitude $\Theta(p^2)$ in
(\ref{twovec}) is linear in the momentum squared, the numerator in the
scattering amplitude (\ref{scatt}) is not. However, the change in  the
numerator moving from $t=\mrhoq$ to $t<0$ is not negligible,
a result which was first claimed by Goldmann et al. \cite{gold92}.\\[5mm]
{\large {\bf Acknowledgements}}\\[2mm]
I thank A.Bramon, J.Gasser and H.Genz for helpful discussions, and
U.B\"urgi for providing me with numerical results of his diploma thesis.

\end{document}